\documentclass[12pt,preprint]{aastex}
\usepackage{natbib}
\bibpunct{(}{)}{,}{a}{}{,}
\usepackage{color}
\usepackage[colorlinks=true,citecolor=blue,urlcolor=green]{hyperref}
\usepackage[modulo]{lineno}

\def\teff{T_{\rm eff}}
\def\feh{\rm[Fe/H]}
\def\ali{A({\rm Li})}
\newcommand\nion[2]{#1\,\lowercase{{\sc #2}}}
\def\logg{\log~g}

\def\cfe{\rm[C/Fe]}
\def\nfe{\rm[N/Fe]}

\def\ciso{$^{12}$C/$^{13}$C}
\def\kmsec{\mbox{km~s$^{\rm -1}$}}

\shorttitle{Li-rich giants as seen by {\it Kepler}}
\shortauthors{Silva Aguirre et al.}

\begin{document}

\title{Old puzzle, new insights: a lithium rich giant quietly burning helium in its core\footnote{Based on observations made with the Nordic Optical Telescope, operated by the Nordic Optical Telescope Scientific Association at the Observatorio del Roque de los Muchachos, La Palma, Spain, of the Instituto de Astrof\'isica de Canarias.}}
\author{V.~Silva Aguirre\altaffilmark{2}, G.~R.~Ruchti\altaffilmark{3}, S.~Hekker\altaffilmark{4,5}, S.~Cassisi\altaffilmark{6}, J.~Christensen-Dalsgaard\altaffilmark{2}, A.~Datta\altaffilmark{7}, A.~Jendreieck\altaffilmark{8}, J.~Jessen-Hansen\altaffilmark{2}, A.~Mazumdar\altaffilmark{9}, B.~Mosser\altaffilmark{10}, D.~Stello\altaffilmark{11}, P.~G.~Beck\altaffilmark{12}, and J.~de~Ridder\altaffilmark{12}}
\altaffiltext{2}{Stellar Astrophysics Centre, Department of Physics and Astronomy, Aarhus University, Ny Munkegade 120, DK-8000 Aarhus C, Denmark}
\altaffiltext{3}{Lund Observatory, Department of Astronomy and Theoretical Physics, Box 43, SE-221 00 Lund, Sweden}
\altaffiltext{4}{Max-Planck-Institut f\"{u}r Sonnensystemforschung, Justus-von-Liebig-Weg 3, 37077, G\"{o}ttingen, Germany}
\altaffiltext{5}{Astronomical Institute `Anton Pannekoek', University of Amsterdam, Science Park 904, 1098 HX Amsterdam, the Netherlands}
\altaffiltext{6}{INAF-Astronomical Observatory of Teramo, Via M. Maggini sn, I-64100 Teramo, Italy}
\altaffiltext{7}{Department of Physics and Meteorology, Indian Institute of Technology, Kharagpur - 721302, India}
\altaffiltext{8}{Max-Planck-Institut f\"{u}r Astrophysics, Karl Schwarzschild Strasse 1, 85748, Garching, Germany}
\altaffiltext{9}{Homi Bhabha Centre for Science Education, TIFR, V. N. Purav Marg, Mankhurd, Mumbai 400088, India}
\altaffiltext{10}{LESIA Ð Observatoire de Paris, CNRS, Universit\'e Pierre et Marie Curie, Universit\'e Denis Diderot, 92195 Meudon Cedex, France}
\altaffiltext{11}{Sydney Institute for Astronomy, School of Physics, University of Sydney, NSW 2006, Australia}
\altaffiltext{12}{Instituut voor Sterrenkunde, KU Leuven, Celestijnenlaan 200D, B-3001 Leuven, Belgium}
\begin{abstract}
About 1\% of giant stars have been shown to have large surface Li abundances, which is unexpected according to standard stellar evolution models. Several scenarios for lithium production have been proposed, but it is still unclear why these Li-rich giants exist. A missing piece in this puzzle is the knowledge of the exact stage of evolution of these stars. Using low-and-high-resolution spectroscopic observations, we have undertaken a survey of lithium-rich giants in the {\it Kepler} field. In this letter, we report the finding of the first confirmed Li-rich core-helium-burning giant, as revealed by asteroseismic analysis. The evolutionary timescales constrained by its mass suggest that Li-production most likely took place through non-canonical mixing at the RGB-tip, possibly during the helium flash.
\end{abstract}


\section{Introduction}\label{sec-int}
\linenumbers
Lithium nuclei are readily destroyed via proton-capture when they are exposed to temperatures exceeding $\sim2.6\mathsf{x}10^6$~K. As a star leaves the main-sequence phase, its surface Li abundance is expected to decrease due to the inward penetration of the convective envelope called the first dredge-up \citep[FDU, e.g.,][]{salaris02}. This process carries material from the surface to hotter interior regions where lithium is burned, depleting its amount compared to the initial value \citep{iben67}. As a consequence, the expected canonical surface lithium abundance of a 1.5~M$_\sun$ star at the end of the FDU is $\ali\sim1.5\footnote{\ali\,=\,$\log\left(\frac{N(\mathrm{Li})}{N(\mathrm{H})}\right)+12$}$ \citep[e.g.,][]{palmerini11}.

Nevertheless, about 1\% of giant stars show an unusual enhancement in their surface Li abundance \citep[cf.,][]{brown89}, an occurrence that challenges standard stellar evolution models. A variety of scenarios have been proposed to explain this phenomenon \citep[e.g.,][]{sackmann99,boothroyd99,romano99,cb00}, commonly featuring Li enrichment during the asymptotic-giant branch (AGB) phase or at the luminosity function bump in the red-giant branch (RGB). Evolutionary classifications of many known Li-rich giants, primarily based on their location in the Hertzsprung-Russell diagram (HRD), have been found consistent with this assumption.

However, observations have revealed the existence of Li-rich giants at different luminosities along the RGB \citep[e.g.,][]{monaco11}, whereas any lithium enhancement taking place at the bump phase is expected to be depleted by the time the star reaches the RGB-tip. Moreover, \citet{kumar11} found Li-rich giants having atmospheric properties in agreement with them being in the clump phase \citep[see also][]{martell13}. Finding Li-rich giants in the core-helium burning phase is evidence of a different enrichment scenario that could be triggered by non-canonical mixing at the RGB-tip (e.g., during core helium ignition) or at the clump phase itself.

If we aim at unveiling the processes that enhance lithium in red giants, detailed elemental abundances from high-resolution spectroscopy and a clear classification of the evolutionary phase of these stars is absolutely critical. This is now possible thanks to asteroseismic observations from space-borne missions, which allow discrimination between red giants in the hydrogen-shell or core-helium burning phases \citep{bedding11a}. A lithium-rich giant below the RGB-bump luminosity has recently been found by \citet{andtwa13} in the {\it Kepler} field. In this Letter, we report the discovery of the first confirmed Li-rich clump red-giant star, KIC~5000307. We determine the combined spectroscopic and asteroseismic properties of this target, and use these data to investigate the mixing and nucleosynthesis processes that could originate its Li-enhancement.
\section{Candidate selection and observations}
As Li-rich giants are extremely rare, large spectroscopic surveys are ideal for their identification. The Large Sky Area Multi-Object Fiber Spectroscopic Telescope \citep[LAMOST;][]{cui12} is well-suited for the discovery of these rare objects, since its low-resolution ($R\sim2,000$) spectra has a wavelength coverage containing the 6708~\AA{} Li line. Using synthetic spectra, degraded to the resolution of LAMOST, we found that a Li-6708 line with an equivalent width $W_{\lambda}>200~$m\AA{} is detectable. We used the LAMOST observations of the {\it Kepler} field to search for potential Li-rich candidates. To confirm the Li abundance and derive more accurate stellar parameters, we obtained high-resolution spectroscopic observations using the Fibre-fed Echelle Spectrograph (FIES) on the Nordic Optical Telescope (NOT) located at La Palma Observatory in the Canary Islands, Spain. We set FIES to deliver a resolving power of $R\sim46,000$ with a spectral coverage of 3700~\AA{} to 7300~\AA~(all spectra were reduced using the FIEStool\footnote{http://www.not.iac.es/instruments/fies/fiestool/FIEStool.html} reduction software).
\section{Spectroscopic analysis}\label{sec_spec}
All parameters of KIC~5000307 derived directly from spectroscopic and asteroseismic observations, as well as those determined in combination with stellar models, are listed in Table~\ref{star_data}.
\subsection{Atmospheric parameters}\label{ssec_atmos}
The spectroscopic stellar parameters (effective temperature, surface gravity, and metallicity) were determined following the iterative methodology described in \citet{ruchti13}, which uses on-the-fly non-local thermodynamic equilibrium (NLTE) corrections to \nion{Fe}{I} lines. Briefly, the spectral analysis was performed with the MOOG program \citep{sneden73}, using one-dimensional, plane-parallel Kurucz model atmospheres \citep[][and references therein]{castelli03}, which are computed under the assumption of local thermodynamic and hydrostatic equilibrium. The effective temperature, $\teff$, was derived from the wings of the Balmer lines through profile fits to H$\alpha$ and H$\beta$. The microturbulence was found by minimizing the slope of the relationship between the NLTE-corrected abundance of iron from \nion{Fe}{I} lines and the reduced $W_{\lambda}$. The surface gravity, $\logg$, and metallicity, $\feh$, were then derived by minimizing the difference between the abundance of iron from the NLTE-corrected \nion{Fe}{I} lines and that from the \nion{Fe}{II} lines. We note that the spectroscopic $\logg$ value agrees well with the one derived using asteroseismology (see Section~\ref{sec_ast} below).
\subsection{Lithium and CN abundances}\label{ssec_li}
The $^7$Li abundance was determined following the methodology described in \citet{ruchti11}. We measured an equivalent width of 250~m\AA{} for the 6708~\AA~Li line. Using this value we derived a Li abundance of $\ali=2.80$ in LTE and $\ali=2.71$ in NLTE, computed following the NLTE corrections of \citet{lind09a}. The [C/Fe] and [N/Fe] ratios, as well as the $^{12}$C$/^{13}$C isotopic ratio, were derived from spectral synthesis of CH and CN lines (B. Plez 2011, priv. comm.) using MOOG under molecular equilibrium. Details of the oscillator strengths and dissociation energies can be found in \citet{ruchti11}.

\begin{figure*}
\includegraphics[width=\linewidth,height=19cm]{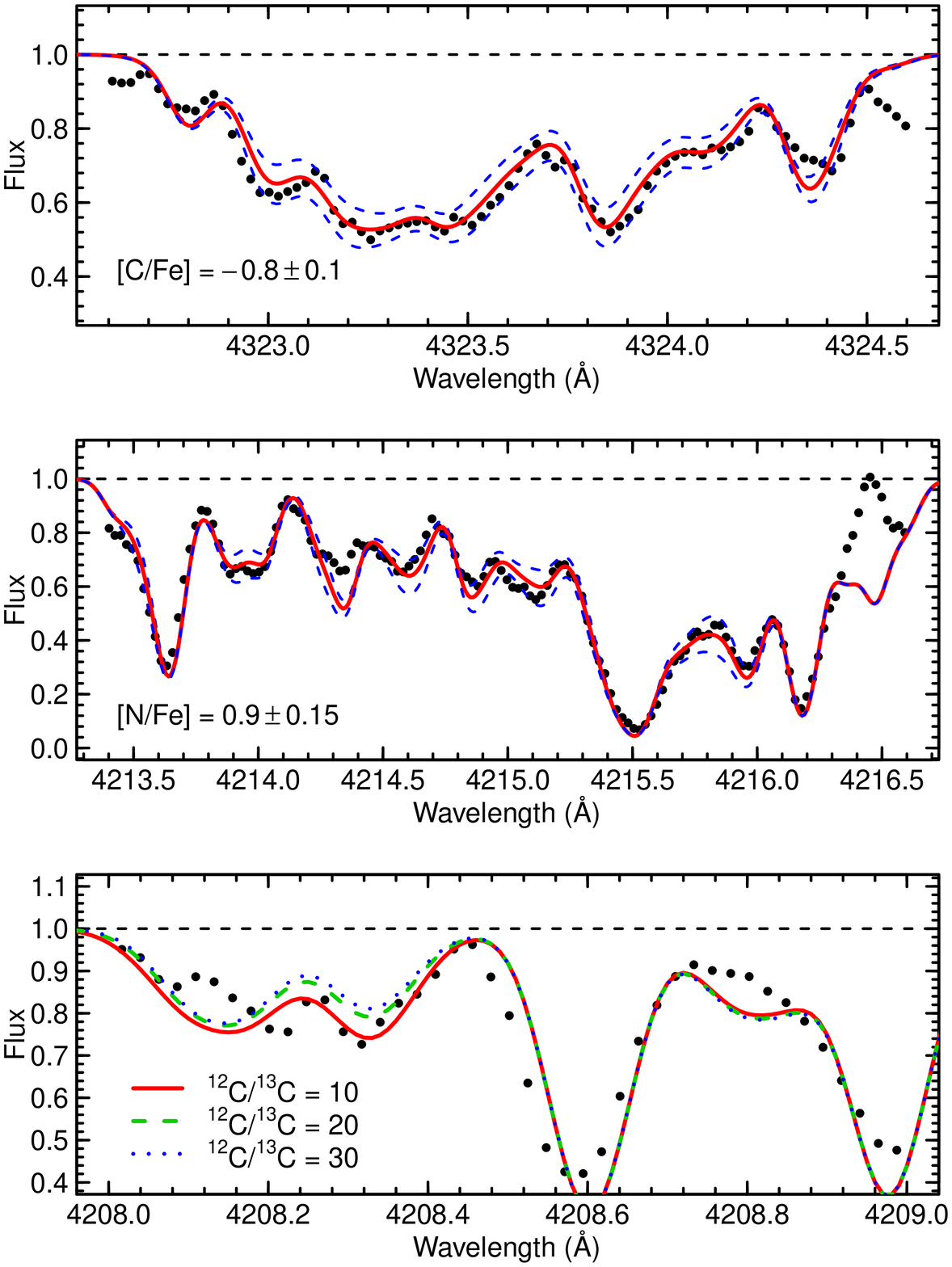}
\caption{Resultant spectroscopic syntheses. \textit{Top:} Fit to the CH G-band to estimate the carbon abundance. Black points represent the observed data. The solid, red curve shows the best fit, while the blue-dashed curves indicate the error on this fit. \textit{Middle:} Synthesis of the CN band around 4215~\AA{} to estimate [N/Fe]. The curves and points are the same as in the top plot. \textit{Bottom:} synthesis of the 4208.3~\AA{} CN feature. Three values of the \ciso~ratio equal to 10, 20, and 30 are shown as a red-solid, green-dashed, and blue-dotted curve, respectively.}\label{fig:abund}
\end{figure*}

The syntheses are shown in Fig.~\ref{fig:abund}. The resultant abundances indicate that the star is deficient in carbon, ${\rm[C/Fe]}=-0.8\pm0.1$, while enhanced in nitrogen, ${\rm[N/Fe]}=+0.9\pm0.15$. Features sensitive to the \ciso~ratio are all very weak. We can thus only place a limit of \ciso~$<$~20 using the CN feature at 4208.3~\AA{}. Unfortunately, all oxygen lines are affected by telluric emission which prevented us from determining the oxygen abundance.
\section{Asteroseismic analysis}\label{sec_ast}
Stars with outer convective envelopes show stochastically excited oscillations that travel across their interiors, called solar-like oscillations. Observing the frequency of these pulsations yields immediate information about the size of the cavity where these waves propagate, providing stringent constraints on its physical properties. Low- and intermediate-mass red giants exhibit this type of pulsations, clearly visible in the data obtained by the CoRoT and {\it Kepler} missions \citep{deridder09,huber11}.

\begin{figure}
\includegraphics[width=\linewidth]{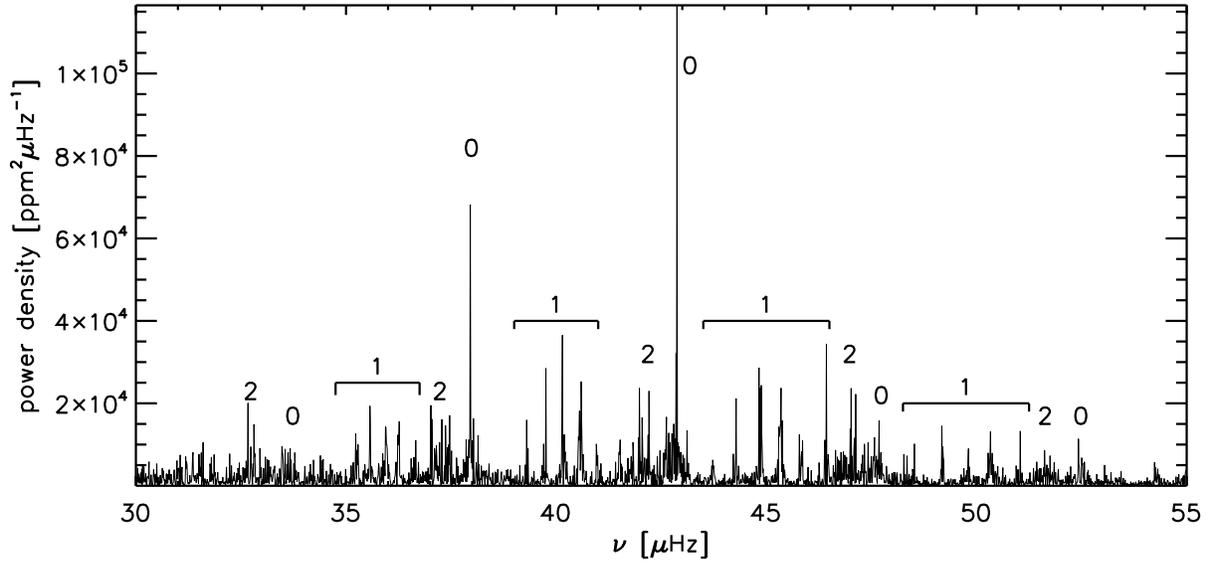}
\caption{Power spectrum of KIC~5000307, showing the oscillation frequencies. Numbers identify peaks of equal angular degree $\ell$, with the $\ell=1$ mixed modes marked by horizontal lines.}\label{fig:pow}
\end{figure}

For the asteroseismic analysis we have used nearly three years of \textit{Kepler} data (Q0-Q12). The Kepler light curve has been extracted using the pixel data following the methods described in Mathur et al. (2014, in preparation) and corrected following \citet{garcia11}. The power spectrum of KIC~5000307 is shown in Fig.~\ref{fig:pow}, where the angular degree $\ell$ describing the geometrical component of each oscillation mode is labeled. Modes of the same degree and consecutive order $n$ are approximately equally spaced in frequency, as can be seen for the $\ell=0$ case in Fig.~\ref{fig:pow}. This quantity is called the large frequency separation $\Delta\nu=\nu_{\ell,n}-\nu_{\ell,n-1}$. The overall power spectrum shows a Gaussian-shaped envelope \citep[e.g.,][]{chaplin13}, where the frequency of maximum oscillation power is known as $\nu_{\rm max}$. These two global oscillation parameters, $\Delta\nu$ and $\nu_{\rm max}$, have been determined using the method described by \citet{hekker10}.

Solar-like oscillations originate from two types of standing waves: those whose restoring force is the pressure gradient (called p-modes) and those where buoyancy acts as the restoring quantity (called g-modes). In main-sequence stars only pure p-modes reach observable amplitudes at the stellar surface, showing a very regular pattern for every angular degree \citep[e.g.,][]{bedding11b}. When a star leaves the main sequence, hydrogen burning ceases in the center and is restricted to a thin shell outside the inert helium core. The star's envelope expands and its core contracts, inducing a large increase in the buoyancy frequency in the centre. This produces coupling between the cavities where p-modes and g-modes reside, resulting in mixed non-radial ($\ell\ge1$) modes whose pulsation frequencies behaving like p-modes in the outer envelope and g-modes in the deep interior \citep[see e.g.,][for a review]{chaplin13}. 

Red giants show a very dense spectrum of modes of mixed character, as seen in Fig.~\ref{fig:pow} for the dipole ($\ell=1$) modes. The underlying g-modes are approximately equally spaced in period, and measuring this separation provides strong constraints on the evolutionary stage of the star \citep[see Fig.~3 in][]{bedding11a}. The period spacing of dipole modes ($\Delta\Pi_1$) has been determined by two independently developed methods, those of \citet{mosser12a} and Datta et al. (in preparation). Briefly, the latter method uses the empirical Lorentzian variation of the observed period spacing around an underlying p-mode to determine the $\Delta\Pi_1$, or the vertical stacking in a period-\'echelle diagram. \citet{mosser12a} on the other hand uses an asymptotic expansion of mixed modes. Regardless of the method employed, the resulting period spacing values are consistent within the uncertainties quoted in Table~\ref{star_data}.

\begin{figure}
\includegraphics[width=\linewidth]{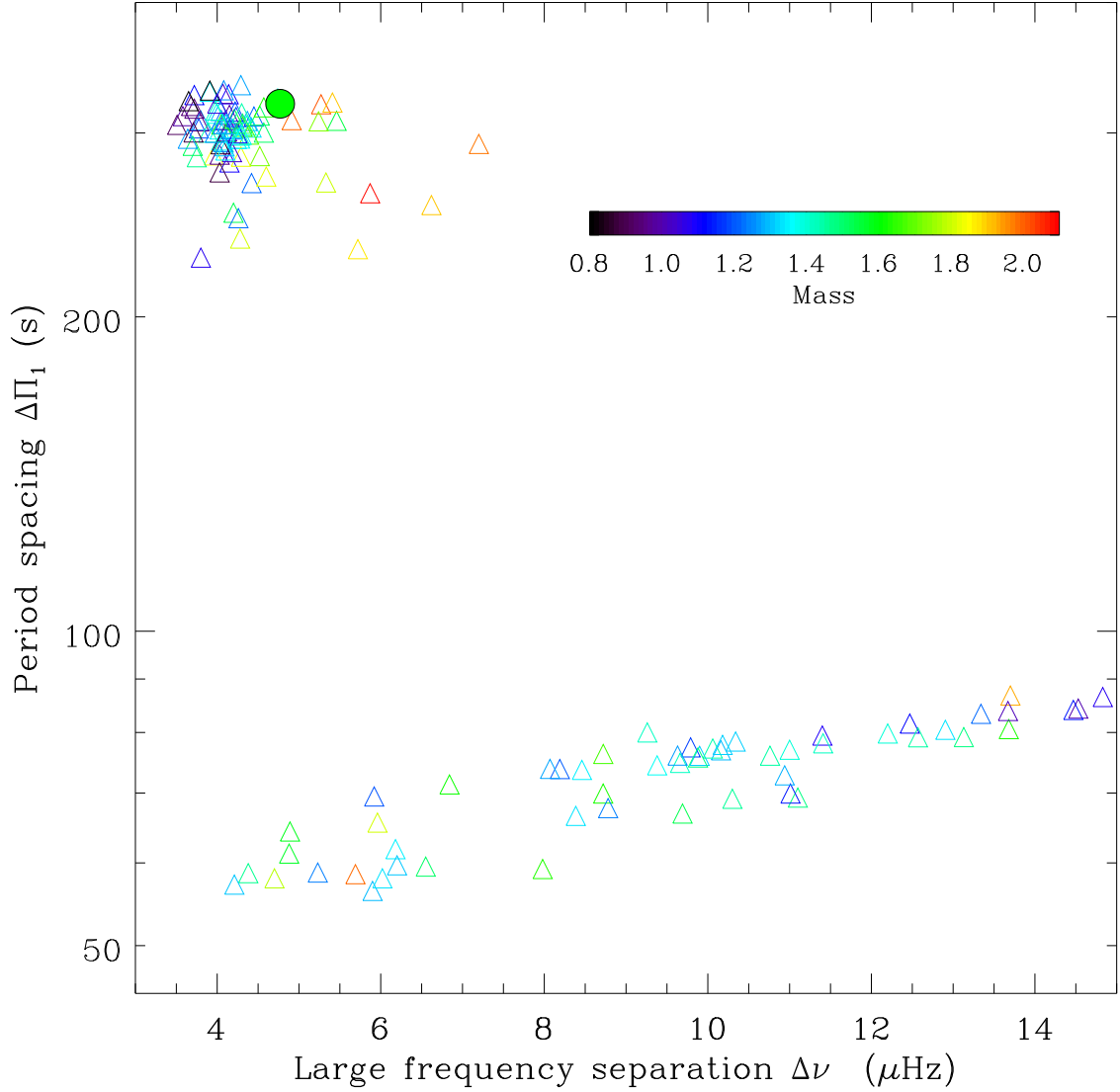}
\caption{Sample of {\it Kepler} stars color coded by mass as determined by \citet{mosser12a}. Stars in the RGB phase of evolution ($\Delta\Pi_1\le100$~s) and the clump phase ($\Delta\Pi_1\ge150$~s) are clearly separated in this diagram. Location of KIC~5000307 is marked with a filled circle.}\label{fig:dnudp}
\end{figure}

In Fig.~\ref{fig:dnudp} we show the $\Delta\nu$ versus $\Delta\Pi_1$ diagram of the {\it Kepler} sample analyzed by \citet{mosser12a}. This figure can be used to effectively discriminate between stars in the ascending red giant branch phase and those burning helium in the core \citep{bedding11a,mosser11}. Our target sits in the region where clump stars are (period spacing values above $\sim$150~s), confirming that this Li-rich giant has ignited helium in its core. Combining this with its value of $\Delta\nu$, a proxy for the stellar mean density and thus very sensitive to the stellar radius, ensures that the star has not yet evolved towards the AGB phase \citep[see][]{stello13}.

To determine the asteroseismic mass, radius, and age of KIC~5000307 we apply a technique known as the grid-based method \citep[e.g.,][]{stello09,silvagui12}. Details on the tracks and equations used to determine the theoretical asteroseismic quantities can be found in section~3 of \citet{silvagui13}. We use BaSTI isochrones \citep{pietrinferni04} including the effects of core overshooting during the main-sequence and semiconvection during the clump phase, as suggested by observations of dipole modes period spacing in {\it Kepler} red giants \citep{montal13}.

The input parameters fed to the grid-based analysis are the spectroscopic temperature and metallicity, and the asteroseismic global quantities $\Delta\nu$ and $\nu_{\rm max}$. The method is applied using the Bayesian scheme described in \citet{serenelli13}. Knowledge of the evolutionary stage of the target is implemented as a prior on the bayesian probabilities that allows a precise determination of the stellar age. The asteroseismic mass of $\sim$1.5~M$_\odot$ is consistent with a star that violently ignited helium in a flash. Our results are based in a set of isochrones not considering mass-loss in the RGB, and to include systematic effects particularly affecting the age estimate we have added in quadrature to the uncertainties the difference in the central values obtained with a set of isochrones using a \citet{reimers75} mass loss rate of $\eta=0.4$.

\citet{mosser12b} have determined mean core rotation periods in red giants observed by the {\it Kepler} satellite. Using modes of mixed character, the authors found that stars ascending the red giant branch slightly increase their core rotation periods, while clearly spinning down in the red clump phase. The rotational splitting $\delta\nu_\mathrm{rot}=50$nHz measured in KIC~5000307, corresponding to a period of $\sim100$ days and common to many other clump stars, is evidence of no particularly fast core rotation \citep[see Figs.~6~and~7 in][]{mosser12b}. Similarly, the predicted mean envelope rotation from the modulation of the rotational splitting is much slower than the mean core rotation \citep[see][]{goupil13}.
\section{Discussion}\label{sec_pro}
\begin{figure}
\includegraphics[width=\linewidth]{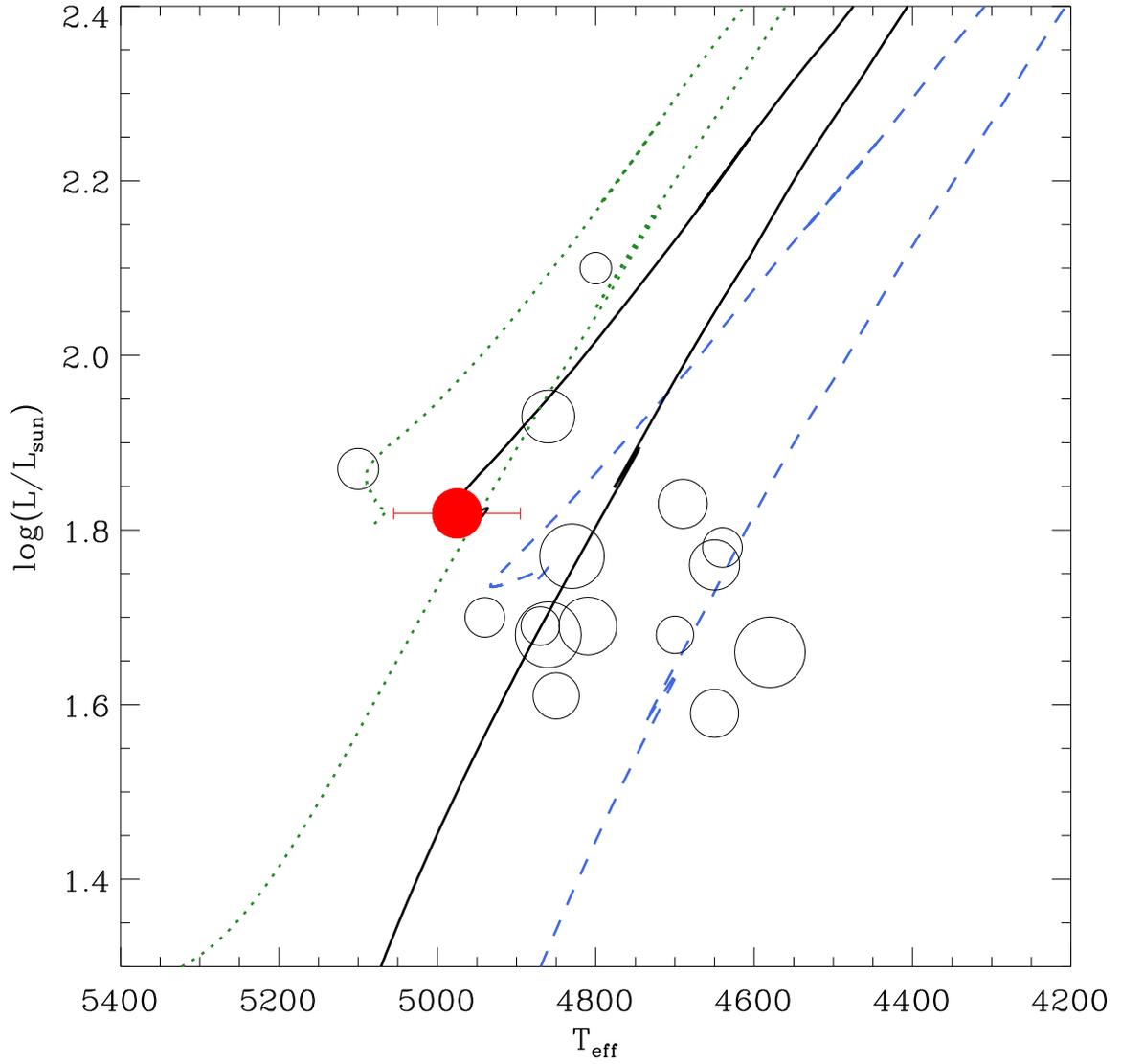}
\caption{Position in the HRD of Li-rich giants reported in \citet{kumar11} (open circles), while the filled symbol depicts our target KIC~5000307. Also shown are BaSTI evolutionary tracks at [Fe/H]=-0.35 for 1~M$_\odot$ (dashed line), 1.5~M$_\odot$ (solid line), and 2~M$_\odot$ (dotted line). The evolution during the helium flash has been removed to avoid clutter. Symbol sizes are scaled according to Li abundances.}\label{fig:compar}
\end{figure}

Combining asteroseismic analysis with classical spectroscopic observations, we have confirmed the first Li-rich giant quiescently burning helium in its core. The obtained mass, radius, age, and period spacing give a consistent picture of a star that has gone through the helium flash. In Fig~\ref{fig:compar} we show its position in the HRD, together with the sample recently identified by \citet{kumar11}. It can be seen that the lack of asteroseismic information could have resulted in KIC~5000307 being mistakenly identified as a $\sim$2~M$_\odot$ star in the RGB phase. Knowledge of the evolutionary stage of the target allows us to test mixing hypotheses for this star.

In order to enhance the surface Li abundance, the $^7{\rm Be}$ isotope (produced in the inner H-burning regions) must be quickly transported by deep circulation to the cooler upper stellar layers before decaying into lithium \citep{cameron71}. Several physical processes have been envisaged as the non-canonical mixing mechanism capable of taking $^7{\rm Be}$ to the stellar surface, such as rapid rotation \citep{drake02}, the interaction with a companion star \citep{denissenkov04}, or magneto-thermohaline mixing \citep{denissenkov09}.

In RGB stars, it is believed that the presence of a steep mean molecular weight gradient left behind by the bottom of the convective envelope at the end of the FDU prevents any extra mixing between the outer convective envelope and the hot layers where H-burning is occurring. However, observations of the carbon isotopic ratio $^{12}$C/$^{13}$C reveal that this quantity decreases below the canonical FDU value of $\sim$25 once the star passes the RGB-bump and evolves towards the RGB-tip \citep[see][]{lind09b,palmerini11}. This is a signature of some non-canonical mixing \citep[such as thermohaline mixing, see][]{charbonnel10} associated with the H-burning shell advancing in mass outwards and crossing the position of maximum convective penetration.

The low [C/Fe] and enhanced [N/Fe] of our target, combined with a likely non-canonical carbon isotopic ratio, describe a consistent picture of additional mixing after the FDU (possibly at the RGB-bump). However, the aforementioned processes also predict a decrease in the Li abundance as the star evolves towards the RGB tip. Thus, even if lithium were produced at the RGB-bump by these mechanisms, it should be depleted by the time the star ignites helium in its core.

\citet{denissenkov12} proposed a mixing scenario in the RGB phase due to fast internal rotation and enhanced mixing across the radiative zone that could explain the lithium abundances and carbon isotopic ratios of the sample stars presented by \citet{kumar11}. Under this prescription, after evolving past the RGB bump the star should zigzag in the HRD towards the position of the clump stars showing $^7$Li enhancement and  $^{12}$C/$^{13}$C depletion in its surface. After this incursion to the clump region, the star resumes its ascent towards the RGB tip before igniting helium. However, our target's $\Delta\Pi_1$ value is evidence of a large convective zone in the stellar center incompatible with the expected core of a RGB star. Moreover, the measured rotational splitting shows no signature of fast core or envelope rotation, ruling out the possibility of KIC~5000307 being an RGB star zigzagging towards the clump region as suggested by \citet{denissenkov12}.

There is no signature of binarity in the {\it Kepler} light curve nor in the spectra's radial velocity, but contamination from a companion star or planet engulfment are possible scenarios for the lithium enrichment of our target. A binary companion going through AGB thermal pulses could produce Li via the Cameron-Fowler mechanism and pollute the surface of our star. However, AGB lithium production is envisaged for masses above $\sim$3~M$_\odot$, thus contamination would have taken place no later than $\sim$300 Myr after the birth of the binary system and would certainly not have survived beyond the FDU in KIC~5000307.

If pollution, on the other hand, took place after the FDU it must have occurred within a narrow time span, namely after the RGB-bump and very close to the RGB tip. Given the target's mass as constrained from asteroseismology, this time frame corresponds to a few tenths of~Myr. It is highly unlikely that pollution from an RGB companion, enhanced in surface lithium abundance and transferring mass to our target, would have occurred in such a short period of time.

A hypothesis first proposed by \citet{delareza96} is that Li-rich giants can be identified using their far-infrared color properties. In this scenario, whatever mechanism responsible for enhancing the surface Li abundance also produces the formation of a circumstellar shell of ejected material, thus affecting the photometric properties of these stars. The only far-IR photometry available for KIC~5000307 comes from the {\it WISE} catalogue \citep{cutri12}, which unfortunately has a large uncertainty in the 22~$\mu$m {\it W4} color and is not conclusive about the presence of IR excess. Further observations in this region of the spectrum would be highly valuable to validate this scenario.

Another possibility is that lithium appeared in the stellar surface as a consequence of non-canonical mixing during helium ignition. Previous studies of the He-flash in very metal-poor low-mass stars suggested that the convective zone produced by the huge energy release of He-burning could penetrate the overlying hydrogen-rich layers \citep[see][and references therein]{scsw:01}. The resulting inward migration of protons (H-injection) into high-temperature regions leads to a H-shell flash. As a consequence, when the convective envelope is deepening and merging with this H-flash driven convective zone, the surface is enriched with a large amount of matter that has been processed in hydrogen fusion.

The occurrence of H-injection has been found in hydrodynamic simulations of the core He-flash by \citet{mocak11} at solar metallicity. Mimicking the hydrodynamical simulations with a hydrostatic evolutionary code, the authors were able to reproduce H-injection in a 1~M$_\odot$ model with the outcome of a strong pollution of the stellar envelope. Interestingly enough, the \citet{mocak11} results show that the surface $^{12}$C/$^{13}$C drops below $\sim$10 while lithium is enriched to a level of $\ali\sim$3.7. However, they predict carbon enrichment in the surface by a factor of 2 or 3, which we do not observe in KIC~5000307.

If non-canonical mixing during the He flash is responsible for the lithium enhancement, Li-rich core-helium burning stars should be mostly concentrated at the early clump stage since any enhancement occurred at the flash is expected to be depleted as the star evolves towards the AGB phase due to the deepening of the outer convection zone. Under that assumption, if KIC~5000307 has only recently gone through the helium flash, remnant processes of that episode could still be adjusting its interior structure and might be visible in the oscillation spectrum. In a future study, we will aim at detailed modeling of this target using individual frequencies of oscillations, as well as extending the sample of Li-rich giants in the {\it Kepler} field to get a more detailed view of Li-enrichment from the RGB to the AGB.
\acknowledgments%
The authors thank the referee for the useful comments that improved the quality of the paper, and M.~Bergemann for providing profile fits of the Balmer lines. Funding for the Stellar Astrophysics Centre is provided by The Danish National Research Foundation (Grant agreement No. DNRF106). The research is supported by the ASTERISK project (ASTERoseismic Investigations with SONG and {\it Kepler}) funded by the European Research Council (Grant agreement No. 267864). Work by G.R.R. is fully supported by Grant No. 2011-5042 from the Swedish Research Council. This work partially used data analyzed under the NASA grant NNX12AE17G and under the European CommunityÕs Seventh Framework Program grant (FP7/2007-2013)/ERC grant agreement n. PROSPERITY and ERC grant agreement nr. 338251 StellarAges.
\clearpage
\begin{center}
\setlength{\tabcolsep}{0.04in}
\begin{deluxetable}{cc}
\tablecolumns{2}
\tabletypesize{\scriptsize}
\tablewidth{0pc}
\tablecaption{Fundamental properties of KIC~5000307}
\tablehead{\colhead{Property} & \colhead{Value}}
\startdata
RA ($^{\circ}$)\tablenotemark{\it a} & 19:13:39.1\\
\noalign{\smallskip}
DEC($^{\circ}$) & +40:11:04.6\\
\noalign{\smallskip}
K$_p$\tablenotemark{\it b} & 11.23\\
\noalign{\smallskip}
FIES Obsdate (yyyymmdd) & 20130429\\
\noalign{\smallskip}
FIES Spectra S/N\tablenotemark{\it c} & 120\\
\hline
\noalign{\smallskip}
$\nu_\mathrm{max}$ ($\mu$Hz)& 42.46$\pm$0.47 \\ 
\noalign{\smallskip}
$\Delta\nu$ ($\mu$Hz) & 4.77$\pm$ 0.04 \\
\noalign{\smallskip}
Period spacing (s) & 319.95$\pm$0.4 \\
\noalign{\smallskip}
Rotational splitting $\delta\nu_\mathrm{rot}$ (nHz) & 50$\pm$10 \\
\noalign{\smallskip}
Mass (M$_\sun$) & $1.536^{+0.059}_{-0.056}$ \\
\noalign{\smallskip}
Radius (R$_\sun$) & $11.011^{+0.217}_{- 0.219}$\\
\noalign{\smallskip}
$\log~g_\mathrm{seis}$ & $2.539^{+0.003}_{-0.004}$\\
\noalign{\smallskip}
$\mathrm{L}/\mathrm{L}_\odot$ & $1.819^{+0.022}_{-0.023}$\\
\noalign{\smallskip}
Age (Myr) & $2082^{+417}_{-414}$\\
\hline
\noalign{\smallskip}
$\feh$\tablenotemark{\it d} & -0.29$\pm$0.05 \\
\noalign{\smallskip}
$\teff$ K & 5000$\pm$70 \\
\noalign{\smallskip}
$\log~g_\mathrm{spec}$ & 2.56$\pm$0.1 \\
\noalign{\smallskip}
v$_t$ (\kmsec) & 1.4$\pm$0.1 \\
\hline
\noalign{\smallskip}
EW Li (6708~\AA)& 250 \\
\noalign{\smallskip}
$\ali$ & 2.80 \\
\noalign{\smallskip}
$\ali_{\rm NLTE}$ & 2.71 \\
\hline
\noalign{\smallskip}
$\cfe$ & -0.80$\pm$0.10 \\
\noalign{\smallskip}
$\nfe$ & 0.90$\pm$0.15 \\
\noalign{\smallskip}
$^{12}$C/$^{13}$C & $<$20 \\
\enddata
\label{star_data}
\tablenotetext{a}{equinox 2000}
\tablenotetext{b}{Kepler magnitude.}
\tablenotetext{c}{Estimated near the Li-6708 line.}
\tablenotetext{d}{Solar abundances adopted from \citet{asplund09}.}
\end{deluxetable}
\end{center}
\end{document}